\documentclass[letterpaper, 10 pt, conference]{ieeeconf}  % Comment this line out if you need a4paper

\IEEEoverridecommandlockouts                              % This command is only needed if 
\usepackage{rotating}

\PassOptionsToPackage{hyphens}{url}\usepackage{hyperref}
\newcommand{\specialcell}[2][c]{\begin{tabular}[#1]{@{}c@{}}#2\end{tabular}}

\usepackage{xcolor}
\usepackage{tikz}
\usepackage{textcomp}
\usepackage{hyperref}

\newcommand\copyrighttext{%
  \footnotesize \textcopyright 2021 IEEE.  Personal use of this material is permitted.  Permission from IEEE must be obtained for all other uses, in any current or future media, including reprinting/republishing this material for advertising or promotional purposes, creating new collective works, for resale or redistribution to servers or lists, or reuse of any copyrighted component of this work in other works.}
\newcommand\copyrightnotice{%
\begin{tikzpicture}[remember picture,overlay]
\node[anchor=south,yshift=10pt] at (current page.south) {\fbox{\parbox{\dimexpr\textwidth-\fboxsep-\fboxrule\relax}{\copyrighttext}}};
\end{tikzpicture}%
}

\title{\LARGE \bf
Artifact Detection and Correction in EEG data: A Review 
}

\author{Sari Sadiya$^{1}$, Tuka Alhanai$^{2*}$, Mohammad M Ghassemi$^{1*}$%<-this % stops a space
\thanks{\scriptsize {$^{1}$ Department of Computer Science, Michigan State University,
               East Lansing, MI}}
\thanks{\scriptsize {$^{2}$ Department of Computer Engineering, 
               New York University,
               Abu Dhabi, UAE}}
\thanks{\scriptsize {$^{*}$ Authors contributed equally to this work.}}
}

\begin{document}

\maketitle
\thispagestyle{empty}
\pagestyle{empty}

%%%%%%%%%%%%%%%%%%%%%%%%%%%%%%%%%%%%%%%%%%%%%%%%%%%%%%%%%%%%%%%%%%%%%%%%%%%%%%%%
\begin{abstract}
Electroencephalography (EEG) has countless applications across many of fields. However, EEG applications are limited by low signal-to-noise ratios. Multiple types of artifacts contribute to the noisiness of EEG, and many techniques have been proposed to detect and correct these artifacts. These techniques range from simply detecting and rejecting artifact ridden segments, to extracting the noise component from the EEG signal. In this paper we review a variety of recent and classical techniques for EEG data artifact detection and correction with a focus on the last half-decade. We compare the strengths and weaknesses of the approaches and conclude with proposed future directions for the field.
\end{abstract}

%%%%%%%%%%%%%%%%%%%%%%%%%%%%%%%%%%%%%%%%%%%%%%%%%%%%%%%%%%%%%%%%%%%%%%%%%%%%%%%%
\section{Introduction}
\copyrightnotice
Electroencephalography (EEG) is a non-invasive, inexpensive, and portable neuro-imaging technology, but the low signal-to-noise ratio of EEG limits its ease of adoption and use for the research and commercial communities alike. The low signal-to-noise ratio of EEG is due, in part, to a variety of artifacts including ocular artifacts from blinks and eye movements and muscle artifacts from movements. While EEG data is affordable to collect, it is challenging to use in practice because artifacts correction is a necessary prerequisite for meaningful use. 

To reduce the human labor associated with EEG experimentation (and the requisite data cleansing) researchers have developed several methods for automated artifact detection. Once an artifact has been detected, the corrupted segment may be discarded but discarding segments introduces discontinuities to the signal that may limit its applications. To circumvent discontinuities, artifact correction techniques may be utilized to "correct" the signal. Implementing effective strategies for artifact detection and correction requires careful review of approaches scattered across the scientific literature. In this review, we highlight the key research contributions in the EEG artifact detection and correction domain over the last 7 years, and identify promising directions for further research and development efforts.

\begin{table*}[t]
\begin{center}
\caption{A breakdown of all the papers reviewed in this work, in chronological order (by year of publication). Type: (D)etection, (C)orrection or (R)emoval. See Section \ref{Methodology} for a breakdown of the different metrics. $\dagger$ marks a new dataset.  $\ddagger$ Data characteristic were not reported by the authors.}
 \begin{tabular}{| c || c | c | c | c | c | c | } 
 \hline
   \textbf{Paper}  & \textbf{Artifact} & \textbf{Type} & \textbf{Datasets} & \textbf{Method} & \textbf{Requirements} &\textbf{Performance} \\ [0.5ex] 
 \hline\hline
 \cite{Shamlo2013} & Blinks & D & \specialcell{4256 trials$^{\dagger}$} & \specialcell{ICA} & \specialcell{A dataset of \\ 3452 ICA scalp-maps} & $>0.80$ AUC\\
\hline
 \cite{Nedelcu_2017} & \specialcell{Blinks \\ Muscle} & D & \specialcell{47752 trails$^{\dagger}$\\ 1955 Blinks \\ 4203 Muscle Mov.} & \specialcell{Supervised learning \\ algorithms} & \specialcell{Requires labeling} & \specialcell{$0.98$ F1} \\
\hline
 \cite{Dutta_2017} & \specialcell{Muscle} & R & \specialcell{$\ddagger$} & \specialcell{Hand Crafted \\ EEMD} & \specialcell{Uses expert knowledge} & \specialcell{$.83$ F1} \\
\hline
\cite{Gilberet_2017} & All & R & \specialcell{$\ddagger$} & \specialcell{LDA, SVM, KNN, ICA} & \specialcell{ Requires labeling } & \specialcell{$<0.50 $ F1} \\
\hline
 \cite{Nejedly_2018} & \specialcell{All} & D & \specialcell{$\ddagger$} & \specialcell{CNN classifier} & \specialcell{Requires labeling} & \specialcell{$0.92$ F1} \\
\hline
  \cite{Somers_2018} & \specialcell{All} & R & \specialcell{2 new datsets\\real and simulated$^{\dagger}$} & \specialcell{MWF} & \specialcell{Requires labeling \\ assumes stationarity} & \specialcell{$6.20$ SNR} \\
\hline
 \cite{Agarwal2019} & Blinks & D & \specialcell{$\dagger$4 new datasets \\ 2350 blinks} & \specialcell{Hand Crafted} & \specialcell{Assumes artifact frequency } & $>0.94$ F1\\
\hline
% \cite{Kohli_2019} & tES & C & \specialcell{new datset \\ 5 participants} & \specialcell{Adaptive filtering} & \specialcell{Signal and artifact are uncorrelated} & \specialcell{NA}\\
%\hline
\cite{Ghosh_2019} & Blinks & R & \specialcell{$\dagger$2000 trials\\ 1000 Blinks} & \specialcell{SVM \\ Autoencoder} & \specialcell{Requires labeling} & \specialcell{$>0.98$ F1\\ $.024$ RMSE}\\
\hline
\cite{PionTonachini_2019} & \specialcell{Blinks, \\ Muscle, Heart \\ Line, Channel} & D & \specialcell{$\dagger$6352 subjects} & \specialcell{ICA \\ CNN classifier} & \specialcell{Labeling of ICA components} & \specialcell{$0.80$ F1 \\ (multi-class)} \\
\hline
\cite{Blum_2019} & Blinks & R & \specialcell{$\ddagger$2 new dataset\\ simulated and real} & \specialcell{ICA with ASR} & \specialcell{Labeling of ICA components} & \specialcell{Downstream \\ ERP recognition} \\ 
\hline
 \cite{Sadiya2020} & All & R & \specialcell{$\dagger$2 new datasets\\ 4578, 4569 trails \\ 628, 570 artifacts} & \specialcell{Classical classifiers \\ and Autoencoder} & \specialcell{Assumes artifacts \\ are uncommon} & \specialcell{ $0.54$ F1, $0.45$ Kappa \\ Downstream classification } \\
\hline
 \cite{Phadikar_2020} & Blinks & R & \specialcell{modified dataset \\ EEGLAB data \\ with simulated blinks} &  \specialcell{ICA, SVM \\ and Autoencoder} & \specialcell{Uncorrelated signal \\ and noise} & \specialcell{$0.97$ F1 \\ $0.04$ NMSE } \\
\hline
 \cite{zhang2020eegdenoisenet} & \specialcell{Blinks \\ Muscle}  & C & \specialcell{modified dataset \\ with simulated artifact} &  \specialcell{Autoencoder} & \specialcell{Simulates only \\ specific artifacts} & \specialcell{$0.56$ RRMSE} \\
\hline
\end{tabular}
\small
\label{Literature_review:tab}
\end{center}
\vspace{-5mm} %reduce dead space
\end{table*}

%  \cite{Delorme_2007} & \specialcell{Go/Nogo visual \\ Categorization} & \specialcell{Simulated blinks \\ simulated muscle artifacts \\ simulated electrical shift \\ simulated white noise \\ simulated linear trends} & NA & d &  & \specialcell{Extreme Values \\ Linear Trends \\ Data Improbability \\ Kurtosis \\ Spectral Patterns} & \specialcell{ICA (Infomax) \\ ICA (SOBI) \\ ICA (FastICA)} \\
\section{Definition of Artifact} \label{Artifacts}
For the EEG community, an ``artifact'' refers to a diverse set of signal distortions that span spatial, frequency and temporal scales \cite{Louis_2016}. While different taxonomies of artifacts have been proposed \cite{Louis_2016}, the exact distinction between signal and artifact is often dependant on the specific purposes of those collecting the data. For instance, muscle artifacts are unwanted in a motor-imagery Brain Computer Interface (BCI) application, but are useful for tasks such as sleep stage identification \cite{Ghassemi2018snooze}. Given the variety of phenomena that could be classified as an artifact for any given EEG use-case, it is not surprising that artifact detection algorithms are narrowly-focused on correcting the intruding artifact in a specific context \cite{Shamlo2013}. Al alternative approach argues that a distortion to an EEG segment is an artifact \textit{if and only if} the distortion negatively impacts the performance of a downstream tasks \cite{Sadiya2020}. 

\section{Scope of Review} \label{Methodology}
This review includes algorithms for artifact detection and correction using EEG data, \textit{alone}. That is, we do not discuss algorithms that rely on external signals (e.g. electrooculography). Furthermore, we exclude research focused on electrode \textit{`pops'} or other spatially localized artifacts as their unique characteristics enable ease of detection by simple unsupervised and self-supervised techniques \cite{SadiyaBIBM2020}. Finally, for the sake of brevity, when a group of papers constitutes a sequence of incremental improvements, we select only the work which presents the accumulation of that line of research \cite{Chang_2020, Blum_2019}. Table \ref{Literature_review:tab} provides an overview of the literature surveyed in this review. 

\subsection{Removal vs. Correction}
This review distinguishes between two approaches: artifact \textit{removal} and artifact \textit{correction}. For an algorithm to perform correction (rather than removal) it must have access to an artifact free version of the EEG waveform to be used as ground truth for correcting an artifact ridden version of that same waveform. Note that this necessitates that artifact correction algorithms are trained on datasets with simulated artifacts (for instance see the data-set proposed by \cite{zhang2020eegdenoisenet}).

\subsection{Metrics}
The performance of artifact detection algorithms are often measured using manually annotated EEG signals. Common metrics to evaluate artifact detection methods include the F1 score, accuracy, sensitivity, specificity, Area Under the Receiver Operator Curve (AUC), and Cohen's Kappa (inter-rater reliability). For the purpose of comparing performance in this review, we standardized these metrics when possible. For instance, if an author did not report the F1 score, we attempted to derive it from the other metrics \cite{Ghosh_2019}. 

For artifact detection, we compare algorithms using several common performance metrics. We note that not all metrics are equally valid for evaluating EEG artifact detection algorithms. The F1 score and accuracy are appropriate for the assessment of tasks with balanced outcome class labels, which is not common in artifact annotation settings; a classifier graded on an unbalanced dataset may achieve a high accuracy but suffer from a high false negative rate.

Artifact correction algorithms are more challenging to assess compared to detection algorithms as (barring simulated data) the ground truth is unknown. When artifacts are simulated, and access to the artifact free waveform is available, metrics such as normalized mean square error (NMSE) and root mean square error (RMSE) are used \cite{Phadikar_2020,zhang2020eegdenoisenet}. When the data is not simulated the same metrics are calculated using artifact free EEG data collected under similar circumstances (i.e. stimuli and task) \cite{Ghosh_2019}. The signal-to-noise ratio (SNR) between clean and noisy EEG post artifact removal is another popular metric \cite{Somers_2018}. Finally, some researchers use the improvement in downstream task performance as a measure of the reconstruction fidelity; for instance, artifact removal was demonstrated to improve stimuli decoding and visual-evoked potentials recognition \cite{Sadiya2020, Blum_2019}.  

\subsection{Datasets}
Table \ref{Literature_review:tab} lists a summary of investigations conducted for the purpose of developing algorithms for artifact detection and correction. We note that investigators typically evaluate their approaches on data they have collected themselves, as opposed to a standard community benchmark dataset; this highlights a larger issue in the EEG research community around data sharing practices. When data is shared, it is often to study a particular downstream task, so to facilitate this end, artifacts are often removed which renders the dataset irrelevant for the purpose of artifact detection research. For papers surveyed in this review, only a few made their datasets publicly available \cite{Nejedly_2018, Agarwal2019, PionTonachini_2019, Sadiya2020}.

%  Beyond the added hurdle of data scarcity, this also renders many results irreplicable. Adopting more rigorous data sharing practices amongst EEG researchers, as well as in the artifact detection research community specifically, is crucial for creating usable algorithms that can be adopted by various disciplines. 

\section{Artifact Detection Methods} \label{Artifact Detection Methods}
Various machine learning and statistical approaches have been applied to the domain of artifact detection. We elaborate on these methods below.

\subsection{Hand Crafted Methods}
The \textit{BLINK} algorithm was tailor-made to detect the specific signal characteristics of artifacts caused by eye blinks. Like many hand crafted methods, this approach performs well for the specific task it was engineered to accomplish, but can not be easily extended, tuned, or adapted to detect other types of artifacts \cite{Agarwal2019}. %not even if the blink waveform differs from what is expected (two local minimas)  

\subsection{Signal Decomposition Methods}
Blind source separation methods, most prominently Independent Component Analysis (ICA), treat EEG as a composite signal; ICA decomposes EEG signals into their constituent signal components from which an expert may identify and remove artifact components. While there are rules-of-thumb to distinguish artifact from signal components (for instance, higher power aggregates in frontal areas of scalp maps for blinks), expert annotation is still often required. One notable exception to this is the work of \textit{Shamlo et al.}, who side-stepped the need for an expert annotator by collecting thousands of scalp maps of blink artifacts to contrast new EEG segments against \cite{Shamlo2013}. 

\subsection{Supervised Approaches}
Supervised classification approaches including Support Vector Machines (SVM), Decision Trees, and K-nearest neighbors (KNN) have be applied for a variety of EEG artifact detection problems. Deep learning and Neural Network methods are a relatively recent development in the field of EEG artifact detection. Multiple recent efforts have applied Convolutional Neural Networks (CNN) to EEG by representing data as an $n\times t$ image of $n$ channels and $t$ samples. \textit{Nejedly et al.} used a CNN in conjunction with fully automated image processing procedures to automatically detect artifacts in intracerebral EEG data \cite{Nejedly_2018}. Transfer learning has also be used to improve the performance of network models previously trained on different datasets \cite{Nejedly_2018}. Ultimately, supervised classifiers have been shown to effectively discriminate artifact from signal segments \cite{Gilberet_2017,Ghosh_2019}, but require annotated artifact data to do so, which is not commonly available for many EEG datasets.  

\subsection{Unsupervised Approaches}
\textit{Sadiya et al.} proposed a general-purpose artifact detection algorithm \cite{Sadiya2020}; their method extracted $58$ different EEG features that are commonly used in EEG research and prognostication, and made the assumption that the frequency of artifacts in the datasets was relatively low. While, this assumption may not always be true (for instance, seizure detection), it is usually valid. The authors benchmarked multiple unsupervised methods. For instance, an auto-encoder was trained to reconstruct EEG waveform segments. Assuming artifact are infrequent, the auto-encoder minimizes the reconstruction error for artifact free trials, hence high reconstruction error is taken as indicative of an outlier EEG segment likely to be an artifact. Their results showed artifact detection rates comparable to the inter-annotator agreement reported in the literature, but as expected, unsupervised algorithms are outperformed by methods tailor-made to detect a given artifact type (Table \ref{Literature_review:tab}).

\subsection{Hybrid Approaches}
Hybrid methods that use deep learning classifiers in conjunction with other methods have shown great promise. \textit{ICLabel} is a recently available artifact rejection plugin for EEGLab\footnote{\url{https://github.com/sccn/ICLabel}} that uses a CNN to label the components of the ICA decomposed waveform \cite{PionTonachini_2019}. The classifier distinguishes between seven different artifact types with a binary accuracy (artifact vs signal) of $0.83$. Like other ICA based algorithms, ICLabel is capable of online artifact rejection.

\section{Artifact Removal and Correction Methods} \label{Artifact Correction Methods}
Detecting and excluding artifact ridden trails allows researchers access to clean data. However, these trials could constitute a non-trivial portion of the collected data, and rejecting them may introduce discontinuities into data that is fundamentally temporal in nature. Recent research efforts have focused on approximating an artifact free version of the affected segment, instead of discarding it all together. It is important to note that all artifact removal methods discussed below are supervised, even when constituting a component of a larger unsupervised pipeline. 

% For example, while \cite{Sadiya2020} implemented a fully unsupervised outlier detection and removal, the removal component itself still had to be trained using data labeled by a preceding component in the pipeline.

\subsection{Signal Decomposition Methods}
As previously stated, ICA decomposes EEG signals into their constituent components from which noise components may be identified. A natural extension of the detection algorithms discussed above is to reconstruct the EEG signal from all but the identified noise components. \textit{Gilbert et al.} trained several classifiers (LDA, SVM, KNN) to distinguish between signal and noise independent components \cite{Gilberet_2017}, and as previously mentioned, \cite{PionTonachini_2019} trained a CNN classifier to distinguish between noise and signal components. Notably, these methods involve some global loss of information when the signal is reconstructed \cite{Phadikar_2020}.

%For all it's strengths,  not without its drawbacks; ICA does not separate between noise and signal perfectly which . 

Another approach to blind source separation is Artifact Subspace Reconstruction (ASR) which learns statistical characteristics of the components resulting from Principal Component Analysis (PCA). While the performance of ASR and ICA based methods are comparable, the former is faster and less computationally demanding, and is therefore more suitable for online artifact correction \cite{Blum_2019}.

Extended Empirical Mode Decomposition (EEMD) has also used for EEG artifact removal \cite{Dutta_2017}. Empirical mode decomposition methods can be used as filters but are not strictly in the same category. EMDs decompose signals into a special class of generating functions that maximizes the signal-to-noise ratio of the reconstruction. While EMDs might appear reminiscent of ICA, the nature of the decomposition is different. ICA decomposes the data for all EEG channels simultaneously, while EMD and the other filtering methods decompose the signal at each channel separately.

\subsection{Filter-based Methods}
In signal processing, filters are basic sequence-to-sequence elements that suppress unwanted temporal phenomenon. The Multi-Channel Wiener Filter (MWF) has been used to great effect in audio and speech processing; Wiener filters use labeled examples to estimate parameters of the signal and noise waveforms such that that noise waveform may be filtered out while the NMSE between a clean signal and its output is minimized. The amount of labeling required to use MWF is minimal and an EEGLab plugin is publicly available \cite{Somers_2018}\footnote{\url{https://github.com/exporl/mwf-artifact-removal}}. MWF assumes stationary of the EEG and noise profiles but to be fair, many simple classifiers make a similar assumption. With sufficient depth, neural encoder-decoder models can learn to correct multiple artifacts drawn from different distributions. 

\subsection{Supervised Approaches}
Artifact removal with neural networks is a recent development that was been made possible with breakthroughs in sequence-to-sequence modeling tasks using encoder-decoder neural network architectures. Since the ground truth is not usually available, researchers use noisy trials as the input sequence to the encoder-decoder model and artifact free trials as the target sequence \cite{Ghosh_2019}. To facilitate work in artifact correction, \textit{EEGdenoiseNet} was recently published as a bench-marked data set of simulated ocular and muscle artifacts \cite{zhang2020eegdenoisenet}. The package provided by the authors allows for the simulation of various artifacts at various signal-to-noise ratios. The authors implemented fully-connect, convolutional, and recurrent neural networks to bench mark the data-set.
%An alternative method reconstructs signals by utilizing the surrounding segments of the detected EEG artifact segment \cite{Sadiya2020}. By training with artifact free trials, the method ensures that the reconstructed signal approximates an artifact free signal.

\subsection{Unsupervised Approaches}
As discussed, \textit{Sadiya et al.} suggested an unsupervised approach for artifact detection. Assuming a low false positive rate, the authors used the trials marked as artifact free to train an CNN to reconstruct EEG segments using surrounding samples. The trained network was then used to reconstruct artifact ridden segments. By training with artifact free trials, the method ensures that the reconstructed signal approximates an artifact free signal. While the artifact removal component itself was supervised the pipeline as a whole does not require any labeling (due to the artifact detection being unsupervised). Note that this same approach could be used with any other supervised artifact removal component such as \cite{Ghosh_2019,Somers_2018}. This approach remains highly limited by the low accuracy of unsupervised artifact detection (Table \ref{Literature_review:tab}).   

\subsection{Hybrid Methods}
\textit{Phadikar et al.} suggested a hybrid model that uses SVMs to detect noise components in the ICA deconstructed signals and a denoising autoencoder to remove artifacts from the ICA components rather than the raw EEG \cite{Phadikar_2020}. By denoising the ICA components, instead of excluding them from the reconstruction all together, the reconstruction was found to be more accurate.

%Recently some researchers suggested using EOG data as an online feedback mechanism for adaptive filtering \cite{Ghanem_2018}. This however requires additional hardware (EOG) and limits applicability of the filters to ocular artifacts.

\section{Conclusion}
%Despite narrowing the scope of our review, we still found a plethora of relevant research using various methods and frameworks from last half decade alone. This speaks to the growing importance of artifact detection and removal techniques as EEG devices become more prevalent in multiple fields. However, as evident from Table \ref{Literature_review:tab}, the research community is in dire need for a standardized metric, database, and terminology. Especially if the end goal of this research community is to produce usable applications that have been validated on multiple datasets using rigorous statistics. This issue is doubtlessly going to become more severe as with the advancements of algorithms for rejection and removal, as well as data simulation. We are encouraged by recent research that make important data sets available and hope that the readers incorporates our suggestions to their future research in this domain. 
In this review, we provide a succinct overview of EEG artifact detection and correction methods, with a focus on the last 5 years of research. We reviewed many more papers than formally discussed in this article; indeed, there has been an increased interest in artifact detection and removal as EEG devices become more prevalent in multiple fields. 

As evident from Table \ref{Literature_review:tab}, the research community is in dire need for a standardized metric, database,and terminology surrounding the EEG artifact detection task, especially if the goal is to produce usable application that will generalize to multiple datasets, and heterogeneous tasks. The more recent entries in Table \ref{Literature_review:tab} imply a  growing popularity of deep learning techniques comes at the expense of traditional approaches and expert knowledge. However, we note that recent papers successfully drew on the rich history and knowledge developed within the EEG preprocessing community to build hybrid approaches that synthesize deep learning, ICA frameworks \cite{Phadikar_2020}, or features borrowed from EEG prognostication \cite{Sadiya2020}. We believe that hybrid frameworks are an interesting future direction of work in this domain and uniquely situated to combine the strengths of multiple approaches that will advance the current state-of-the-art.

% Considering that artifact detection methods are tested against annotated ground truth, we suggest the adoption of measures of inter-annotator agreement such as Cohen's Kappa.   

% Thoughts about future? Particular area?
% Potential in feature - deep learning based hybrid. Draw on the rich history?


%%%%%%%%%%%%%%%%%%%%%%%%%%%%%%%%%%%%%%%%%%%%%%%%%%%%%%%%%%%%%%%%%%%%%%%%%%%%%%%%

%%%%%%%%%%%%%%%%%%%%%%%%%%%%%%%%%%%%%%%%%%%%%%%%%%%%%%%%%%%%%%%%%%%%%%%%%%%%%%%%

%%%%%%%%%%%%%%%%%%%%%%%%%%%%%%%%%%%%%%%%%%%%%%%%%%%%%%%%%%%%%%%%%%%%%%%%%%%%%%%%
%\section*{APPENDIX}


%\section*{ACKNOWLEDGMENT}

%The preferred spelling of the word ÒacknowledgmentÓ in America is without an ÒeÓ after the ÒgÓ. Avoid the stilted expression, ÒOne of us (R. B. G.) thanks . . .Ó  Instead, try ÒR. B. G. thanksÓ. Put sponsor acknowledgments in the unnumbered footnote on the first page.

%%%%%%%%%%%%%%%%%%%%%%%%%%%%%%%%%%%%%%%%%%%%%%%%%%%%%%%%%%%%%%%%%%%%%%%%%%%%%%%%

%References are important to the reader; therefore, each citation must be complete and correct. If at all possible, references should be commonly available publications.

\bibliographystyle{IEEEtran} 
\bibliography{IEEEabrv,IEEEexample}

% decoding a given subject specific task is a solved problem, suttelties become consecquential soon after, 
%potential for RL to help solved individualized problem: system has to tune some weights learning eeg to movement decoding. real time better and better find a clever way to define a reward function to fine tune to become most effective for a specific person.  

\end{document}